\documentclass[twocolumn,twoside,slac_two]{revtex4}
\usepackage{graphicx}
\usepackage{fancyhdr}
\begin{document}

\title{The ASTRI Project: prototype status and future plans for a Cherenkov dual-mirror small-telescope array}

%

\author{S. Vercellone, O. Catalano, M.C. Maccarone}
\affiliation{INAF -- IASF Palermo, Via U. La Malfa 153, I--90146 Palermo, Italy}
\author{F. Di Pierro, P. Vallania}
\affiliation{INAF -- Osservatorio Astrofisico di Torino, Via Osservatorio 20, I--10025 Pino Torinese, Italy}
\author{G. Bonnoli, R. Canestrari, G. Pareschi}
\affiliation{INAF -- Osservatorio Astronomico di Brera, Via E. Bianchi 46, I--23807 Merate, Italy}
\author{P. Caraveo, N. La Palombara, M. Fiorini, L. Stringhetti}
\affiliation{INAF -- IASF Milano, Via E. Bassini 15, I--20133 Milano, Italy}
\author{E. Giro}
\affiliation{INAF -- Osservatorio Astronomico di Padova, Vicolo dell'Osservatorio 5, I--35122 Padova, Italy}
\author{G. Tosti}
\affiliation{Dip. di Fisica, Universit\`{a} degli Studi di Perugia, I--06123 Perugia, Italy}
\author{on behalf of the ASTRI Collaboration}
\affiliation{http://www.brera.inaf.it/astri/}

\begin{abstract}
ASTRI (``Astrofisica con Specchi a Tecnologia Replicante Italiana'') is a flagship project of the Italian Ministry 
of Education, University and Research. Within this framework, INAF is currently developing a wide field of view 
(9.6$^{\circ}$ in diameter) end-to-end prototype of the CTA small-size telescope (SST), devoted to the investigation 
of the energy range from a fraction of TeV up to tens of TeVs, and scheduled to start data acquisition 
in 2014. For the first time, a dual-mirror Schwarzschild-Couder optical design will be adopted on a Cherenkov
telescope, in order to obtain a compact optical configuration. A second challenging, but innovative
technical solution consists of a modular focal surface camera based on Silicon photo-multipliers with a logical pixel 
size of 6.2\,mm $\times$ 6.2\,mm. Here we describe the current status of the project, the expected performance, 
and its possible evolution in terms of an SST mini-array. This CTA-SST precursor, composed of a few SSTs 
and developed in collaboration with CTA international partners, could not only peruse the technological solutions
adopted by ASTRI, but also address a few scientific test cases that are discussed in detail.
\end{abstract}

\maketitle

\thispagestyle{fancy}


\section{THE ASTRI PROJECT AND PROTOTYPE}
ASTRI (``Astrofisica con Specchi a Tecnologia Replicante Italiana") is a flagship project of the Italian Ministry of
Education, University and Research strictly linked to the development of the ambitious Cherenkov Telescope 
Array (CTA,~\cite{2011ExA....32..193A}). CTA plans the construction of many tens of telescopes divided 
in three kinds of configurations, in order to cover the energy range from a tens of GeV (Large Size Telescope, LST), 
to a tens of TeV (Medium Size Telescope, MST), and up to 100 TeV and beyond (Small Size Telescope, SST).
Within this framework, INAF is currently developing an end-to-end prototype of the CTA small-size telescope 
in a dual-mirror configuration (SST-2M) to be tested under field conditions, and scheduled to start data acquisition in 2014.

\begin{figure*}
  \includegraphics[angle=0,width=.3\textheight]{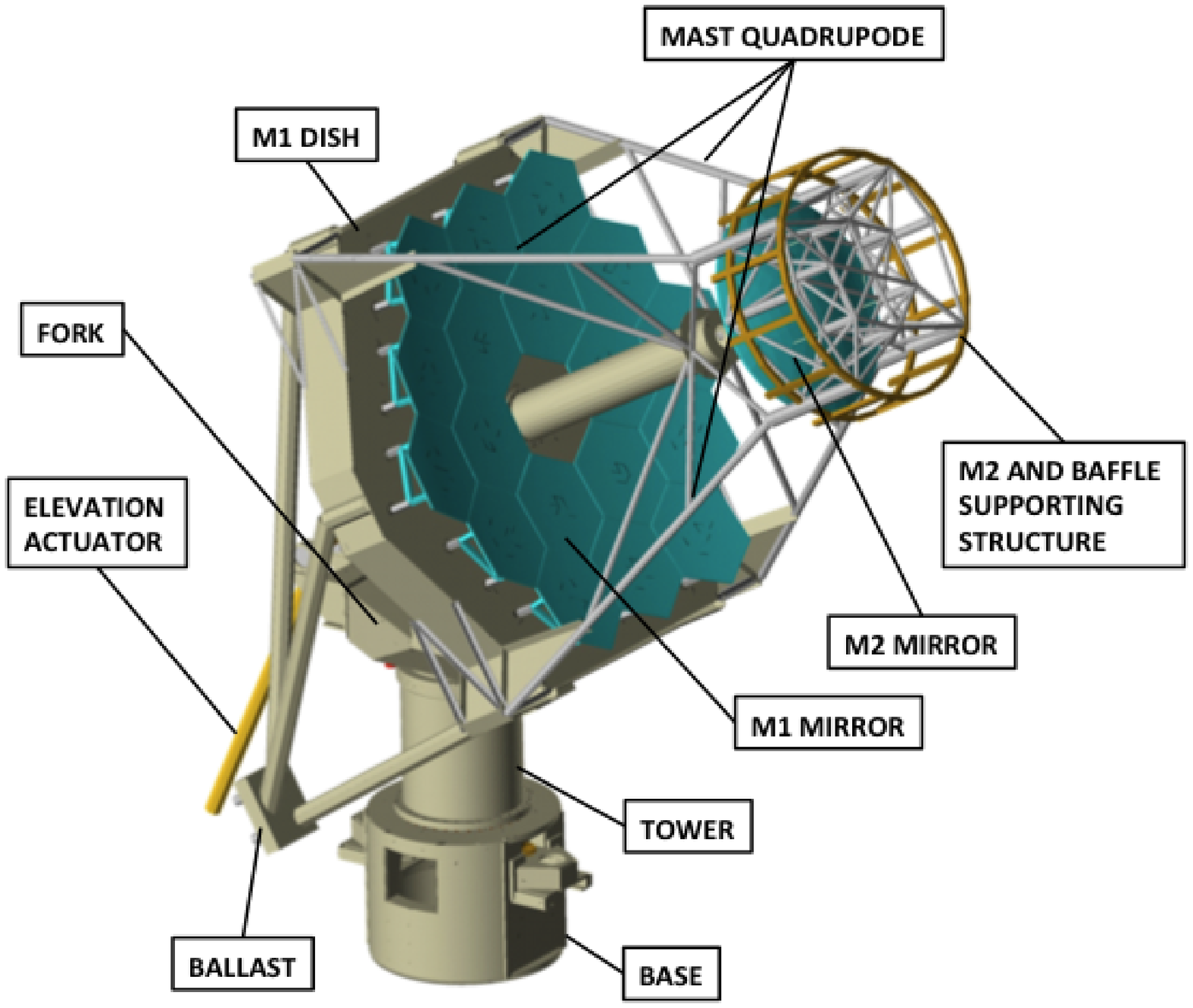}
  \includegraphics[angle=0,width=.3\textheight]{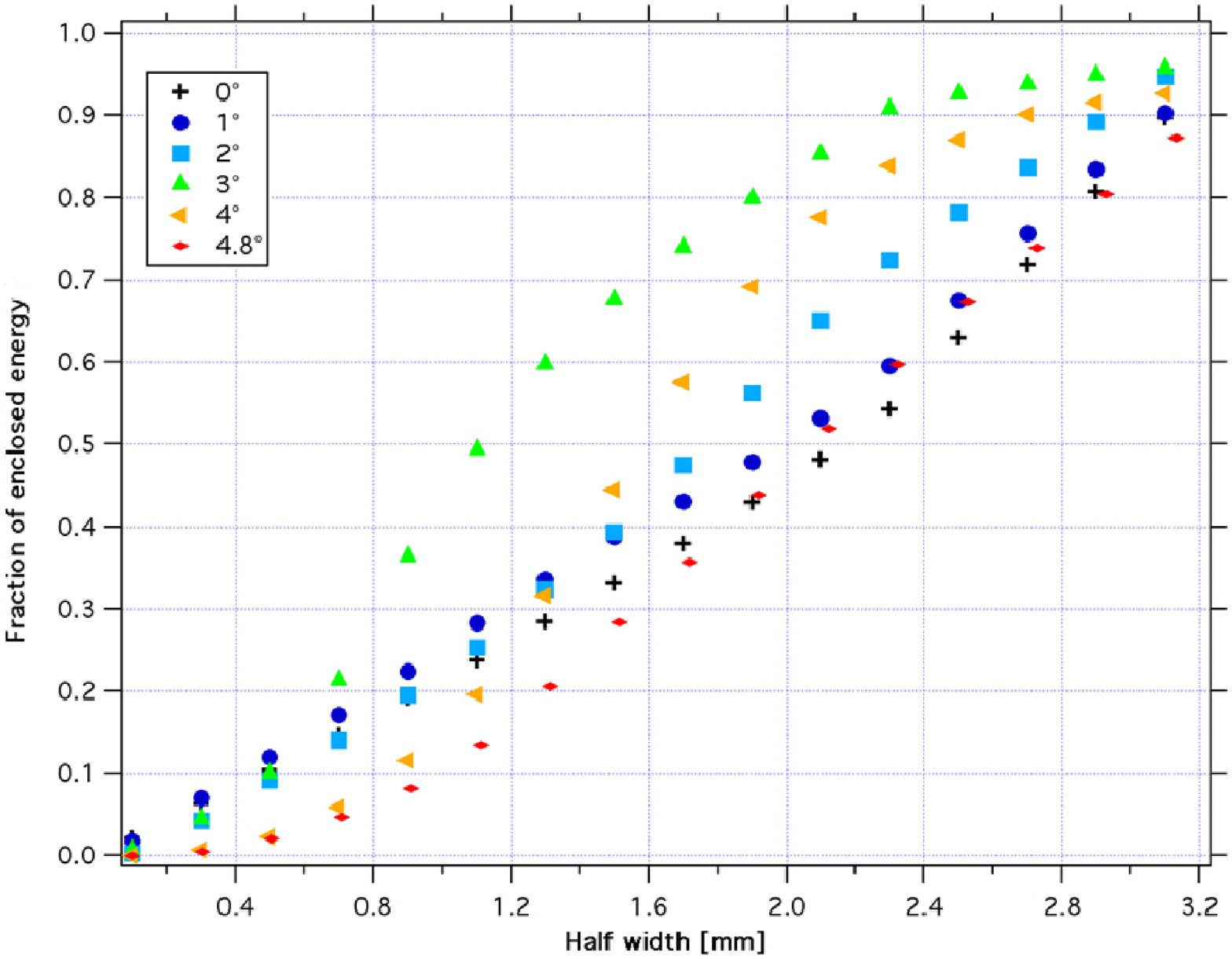}
  \caption{{\it Left panel:} 3-D rendering of the telescope with sub-systems' labels. Courtesy of BCV Progetti 
  and Tomelleri S.r.L. {\it Right panel:} Fraction of the light concentration into half pixel for different incidence angles. 
  }
  \label{fig:ST2_EEW}
\end{figure*}

For the first time, a wide field of view (FoV = $9.6^{\circ}$ in diameter) dual-mirror 
Schwarzschild-Couder (SC,~\cite{2007APh....28...10V}) optical 
design will be adopted on a Cherenkov telescope, in order to obtain a compact (f-number f/0.5) optical configuration 
and equipped with a light and compact camera 
based on Silicon photo-multipliers with a logical pixel size of 6.2\,mm $\times$ 6.2\,mm, corresponding to 
an angular size of $0.17^{\circ}$ (obtained by means of an optimization process among the commercially-available
detectors, the optics performance, and the overall costs of the prototype).
Figure~\ref{fig:ST2_EEW} (left panel) shows the proposed telescope layout, whose mount exploits the classical 
alt-azimuthal configuration, and which is fully compliant with the CTA requirements for the SST array. 
The ASTRI SST--2M prototype will be placed at Serra La Nave, 1735\,m a.s.l. on the Etna Mountain near Catania, 
at the INAF ``M.G. Fracastoro" observing station, and will begin data acquisition in 2014 (\cite{ASTRI-TN-IASFPA-3300-009}).

\section{THE ASTRI DUAL-MIRROR SMALL-SIZE PROTOTYPE}

\subsection{The Optical Design}
The proposed layout~(\cite{ASTRI-IR-OAB-3100-009}) is fully compliant with the CTA requirements
for the SST array. Moreover, our design has been optimized in order to
ensure a light concentration higher than 80\% within the dimension of the pixels over the
entire field of view (Figure~\ref{fig:ST2_EEW}, right panel) and
taking into account the segmentation of the primary mirror (M1) and the dimension and position of the camera.
The telescope design is compact having a 4.3\,m-diameter M1, a 1.8\,m-diameter secondary mirror (M2) 
and a primary-to-secondary distance of 3\,m.
The SC optical design has an f-number f/0.5, a plate scale of 37.5\,mm/$^{\circ}$, a logical 
pixel size of approximately $0.17^{\circ}$ and an equivalent focal length of 2150\,mm. 
Considering 1984 pixels, this setup delivers a FoV of $9.6^{\circ}$ in diameter and
a mean value of the active area of about 6.5\,m$^{2}$, taking into account:
the segmentation of M1,  the obscuration of M2, the obscuration of the camera, the reflectivity of the 
optical surfaces as a function of the wavelength and incident angle, the losses due to the camera's protection 
window and the efficiency of the silicon detectors as function of the incident angles 
(ranging from $25^{\circ}$ to $72^{\circ}$).

\subsection{The Mirrors}
\begin{figure}
  \includegraphics[angle=0,width=.35\textheight]{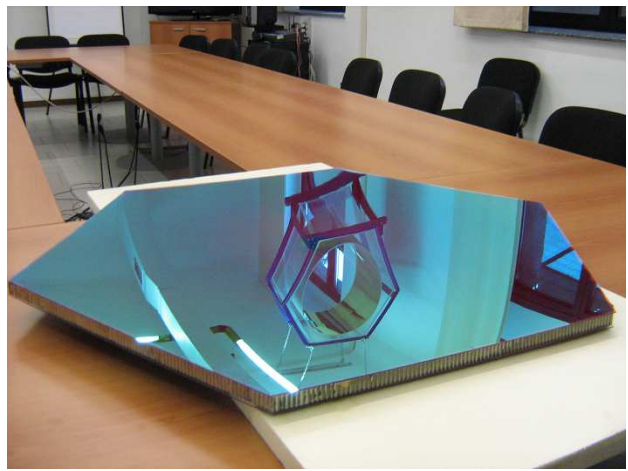}
  \caption{Photographic image of a prototype of one M1 mirror segment.
   }
  \label{fig:MIR}
\end{figure}

The primary mirror M1 is segmented into 18 tiles; the central one is not used because 
completely obstructed by the secondary mirror M2. The segmentation requires three types of segments having 
different surface profiles. Figure~\ref{fig:MIR} shows an image of a prototype of one M1 mirror segment manufactured 
with the glass cold-shaping technology.
The segments have hexagonal shape with an aperture of 849\,mm face-to-face.
Each segment will be equipped with two actuators plus one fixed point for alignment. Only tilt misplacements will 
be corrected. The secondary mirror is monolithic, and has a radius of curvature of 2200\,mm and diameter of 
1800\,mm. M2 will be equipped with three actuators. The third actuator also makes the piston/focus adjustment
 for the entire optical system available.

\subsection{The Telescope Structure}
The optical design  will be implemented by means of a telescope structure 
composed of primary and secondary mirrors cells, a pillar and counterweights, the drives systems and a 
focal surface interface. The telescope mount exploits the classical alt-azimuthal configuration (see
Figure~\ref{fig:ST2_EEW}, left panel).

\subsection{The Camera}\label{subsect:camera}
\begin{figure}
  \includegraphics[angle=0,width=.30\textheight]{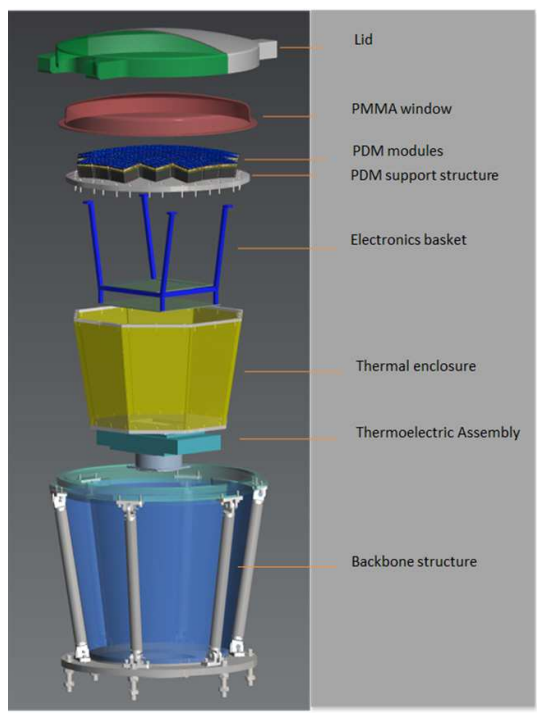}
  \caption{Camera system breakdown at component level.
  }
  \label{fig:CAM_EPS}
\end{figure}

\begin{figure}
  \includegraphics[angle=270,width=.35\textheight]{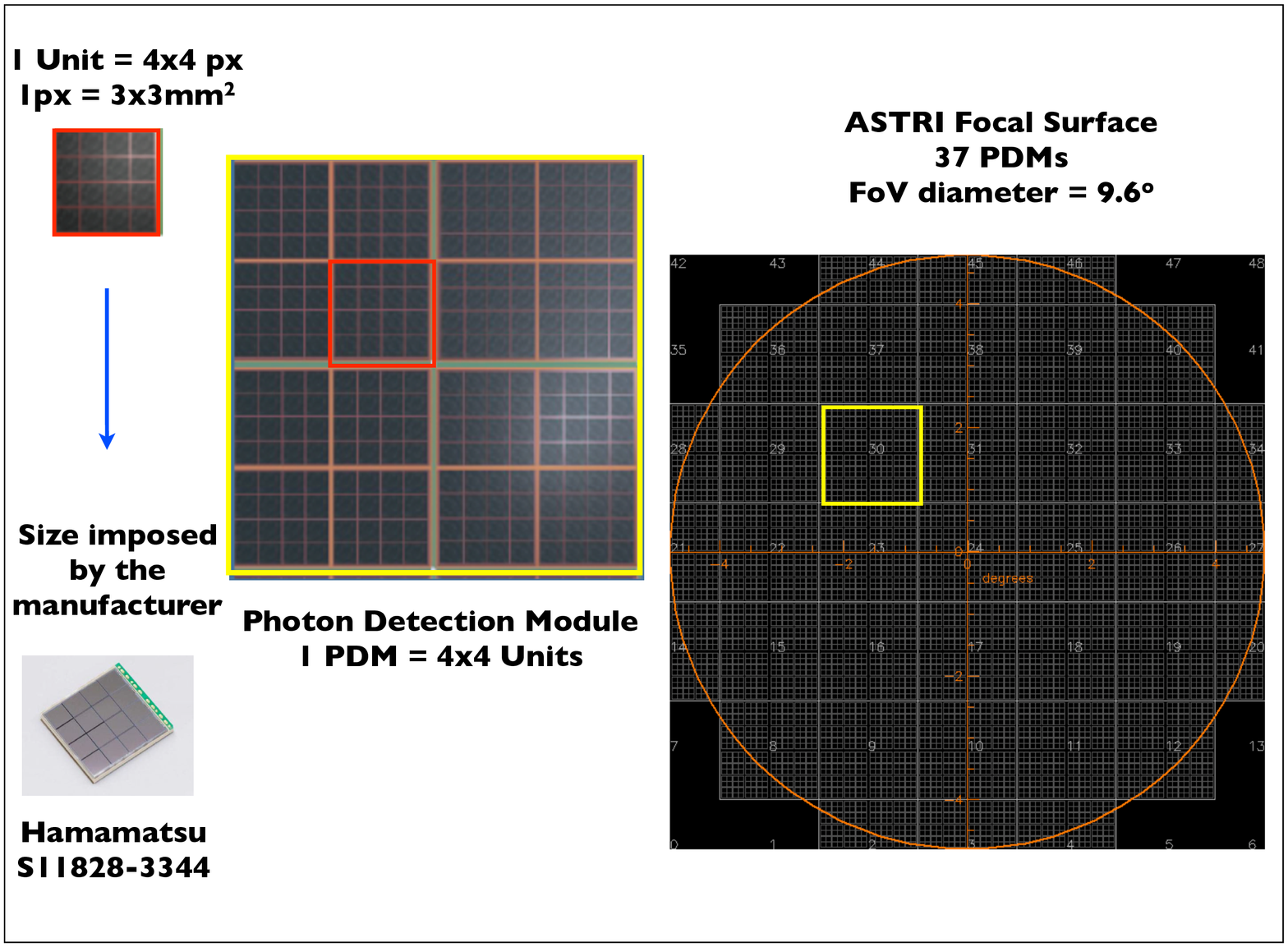}
  \caption{Modular approach adopted to cover with active detectors the entire ASTRI FoV.
  }
  \label{fig:CAM_PS}
\end{figure}

\begin{figure}
  \includegraphics[angle=0,width=.30\textheight]{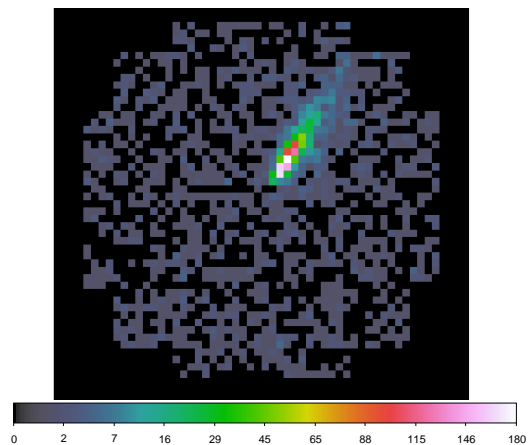}
  \caption{On-axis simulated event on the ASTRI Camera. See Section~\ref{subsect:camera} for details.}
  \label{fig:EVE}
\end{figure}

The SC optical configuration allows us to design a compact and light
camera. Currently, the ASTRI camera has a dimension of about 500\,mm$\times$500\,mm$\times$500\,mm,
including the mechanics and the interface with the telescope structure, for a total weight of about 50\,kg
(see Figure~\ref{fig:CAM_EPS} for a camera system breakdown).
Such small detection surface, in turn, requires a spatial segmentation of a few square millimeters to be compliant
with the imaging resolving angular size. Among the available light sensors that offer
photon detection sensitivity in the 300--700\,nm band, a fast temporal response and a suitable pixel size, 
we selected the Hamamatsu Silicon Photomultiplier (SiPM) S11828-3344M~(\cite{S11828-3344M}).
In order to cover the full 9.6$^\circ$ FoV, we used a modular approach, as shown in Figure~\ref{fig:CAM_PS}.
We call {\it Unit} the physical aggregation (imposed by the manufacturer) of 4$\times$4 pixels 
(3\,mm$\times$3\,mm each pixel). The logical aggregation of 2$\times$2 pixels is called {\it logical pixel}, which
turns out to be of 6.2\,mm$\times$6.2\,mm (0.17$^{\circ}$), while the {\it Photon Detection Module} (PDM) is 
composed of $4\times4$ Units. The ASTRI focal surface, covering the whole FoV, requires 37 PDMs.
The advantage of this design is that each PDM is physically independent of the others, allowing
maintenance of small portions of the camera. To fit the curvature of the focal surface, each PDM is 
appropriately tilted with respect to the optical axis. 
Figure~\ref{fig:EVE} shows an on-axis simulated event for a primary gamma-ray with E=10\,TeV, a core distance
of 142.77\,m and including a night-sky background of $1.9 \times 10^{12}$\,ph\,m$^{-2}$s$^{-1}$sr$^{-1}$,
(about 3\,p.e.\,pixel$^{-1}$). The color-bar shows the number of photo-electrons (p.e.) in each pixel.

\subsection{The Prototype Expected Performance}
\begin{figure}
  \includegraphics[angle=0,width=.35\textheight]{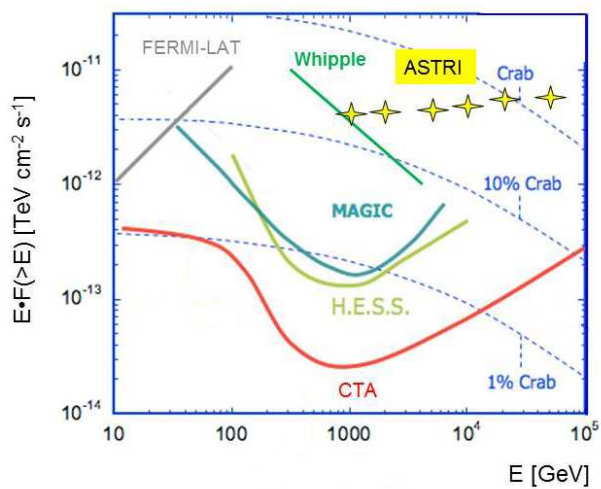}
  \caption{Expected ASTRI SST--2M prototype sensitivity as a function of the energy 
  (yellow stars, computed at 5$\sigma$ and 50\,hr of observation). 
  Adapted from~\cite{ASTRI-MC-IFSITO-5000-001} and reference therein.
  }
  \label{fig:1TS}
\end{figure}

Although ASTRI SST--2M will mainly be a technological prototype, it will perform scientific observations
on the Crab Nebula, MRK~421, and MRK~501. Preliminary calculations~(\cite{ASTRI-MC-IFSITO-5000-001}) show
that in the maximum sensitivity range ($\ge 1$\, TeV) we can detect a flux level of 1~Crab at 5$\sigma$ in a few
hours, while in the energy range $\ge 10$\,TeV a flux level of 1~Crab at 5$\sigma$ can be reached in a few
tens of hours. Figure~\ref{fig:1TS} shows a comparison among the expected ASTRI prototype sensitivity 
as a function of the energy (yellow stars, computed at 5$\sigma$ and 
50\,hr of observation) and those of a few Image Atmospheric Cherenkov Telescope (IACT) ones 
(Whipple, MAGIC, H.E.S.S., CTA) and of large field of view detectors for one-year integration (Fermi-LAT)
Because of their strong flux and spectral variations in the two Markarian sources, estimates of exposures
are more uncertain. In case of large flares, with fluxes up to 5--10 Crab Units, detection could be reached on 
a much shorter time-scale~(\cite{ASTRI-TN-IASFPA-3000-018}), allowing intra-night variability studies.

\section{THE ASTRI SST-2M MINI-ARRAY}\label{sect:miniarray}
\begin{figure}
  \includegraphics[angle=0,width=.35\textheight]{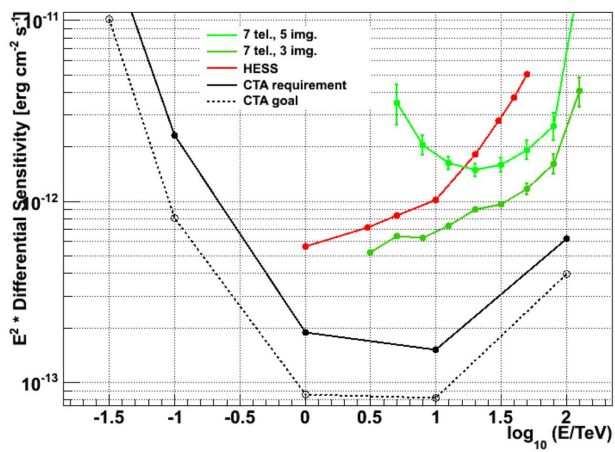}
  \caption{Expected ASTRI SST--2M mini-array sensitivity as a function of the energy (computed at 5$\sigma$ and 
  50\,hr of observation), compared with H.E.S.S. and CTA ones (see Section~\ref{sect:miniarray} for details). 
  From~\cite{ASTRI-MC-IFSITO-5000-005}.
  }
  \label{fig:AR1}
\end{figure}

\begin{figure}
  \includegraphics[angle=0,width=.25\textheight]{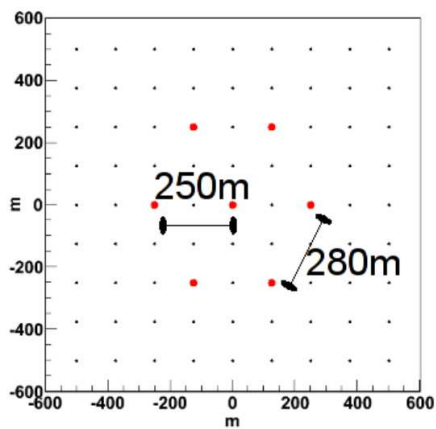}
  \caption{A possible layout for a 7-telescope mini-array configuration (see Section~\ref{sect:miniarray} for details). 
  From~\cite{ASTRI-MC-IFSITO-5000-005}.
  }
  \label{fig:AR2}
\end{figure}

A remarkable improvement in terms of performance could come from the operation, in 2016, of a mini-array, 
composed by a few SST-2M telescopes and to be placed at final CTA Southern Site. 
Preliminary Monte Carlo simulations (\cite{ASTRI-MC-IFSITO-5000-005})
yield an improvement in sensitivity that for 7 telescopes could be a factor 1.5 at 10\,TeV w.r.t. H.E.S.S.,
as shown in Figure~\ref{fig:AR1}.
The ASTRI SST--2M mini-array will be able to study in great detail relatively bright 
(a few $\times 10^{-12}$\,erg\,cm$^{-2}$s$^{-1}$ at 10~TeV) sources with an angular resolution of a few 
arcmin and an energy resolution of about 10--15\,\%. 
The ASTRI SST--2M mini-array sensitivity were calculated taking into account 5 energy bins per decade, a 5$\sigma$
significance, a number of event/energy-bin $\ge 10$, a signal rate $>5$\% w.r.t. the background rate, an
integration time of 50\,hr, a minimum number of images used in the event reconstruction of 3 and 5, respectively,
and for an array configuration as shown in Figure~\ref{fig:AR2}.

Moreover, thanks to the array approach, it will be possible to verify the wide FoV performance to detect very high
energy showers with the core located at a distance up to 500\,m, to compare the mini-array performance with the Monte
Carlo expectations by means of deep observations of few selected targets, and to perform the first CTA science,
with its first solid detections during the first year of operation.
Prominent sources such as extreme blazars (1ES~0229$+$200), nearby well-known BL~Lac objects (MKN~501)
and radio-galaxies, galactic pulsar wind nebulae (Crab Nebula, Vela-X), supernovae remnants
(Vela-junior, RX~J1713.7$-$3946) and microquasars (LS~5039), as well as the Galactic Center can
be observed in a previously unexplored energy range, in order to investigate the electron acceleration and cooling,
relativistic and non relativistic shocks, the search for cosmic-ray (CR) Pevatrons, the study of the CR propagation,
and the impact of the extragalactic background light on the spectra of the sources.
\begin{figure*}[!th]
  \includegraphics[angle=90,width=.40\textheight]{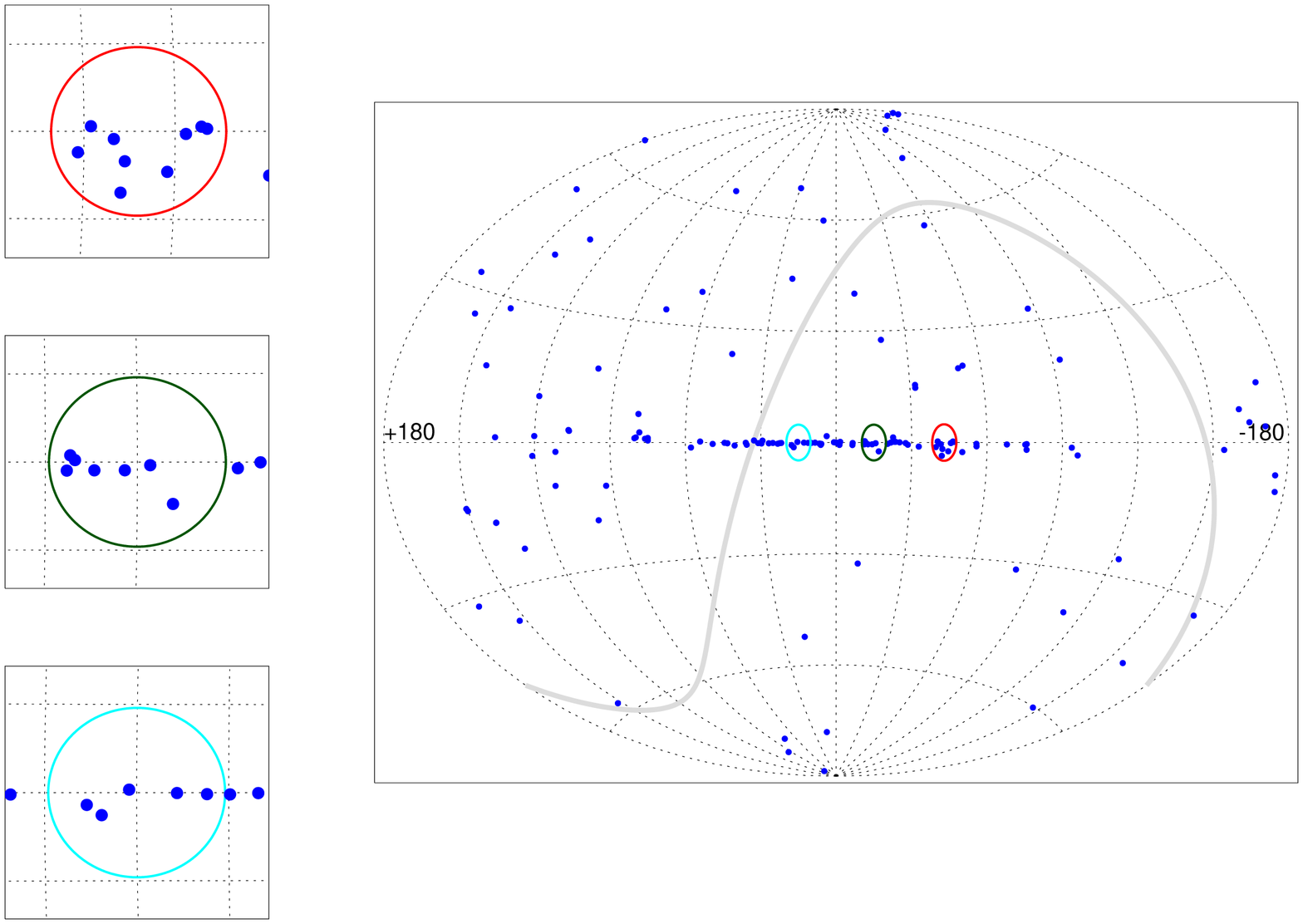}
  \caption{Blue dots are the known TeV sources as listed in the TeVCat Catalogue. The grey line represents the
  Celestial Equator. The red, green and cyan circles are the ASTRI mini-array (optical) field of view. The left panels
  are zooms centered on the ASTRI mini-array pointings.
  }
  \label{fig:FoV}
\end{figure*}
The large field of view of the ASTRI mini-array will allow us to monitor, during a single pointing, a few TeV sources
simultaneously. Figure~\ref{fig:FoV} shows the current TeV sources as listed in the 
TeVCat\footnote{\texttt{http://tevcat.uchicago.edu/}} compilation. Red, green and cyan circles represent the $9.6^{\circ}$ 
(optical) field of view diameter for three possible pointings along the Galactic Plane. 
The grey line represents the Celestial equator.
Although the actual sensitivity will substantially drop for off-axis sources, a few targets can be monitored simultaneously, as 
shown in the three panels on the left. Simultaneous detection of hard and intense Galactic sources could be feasible,
e.g. in the case of Vela--X and Vela--Jr.
\begin{figure}
  \includegraphics[angle=0,width=.35\textheight]{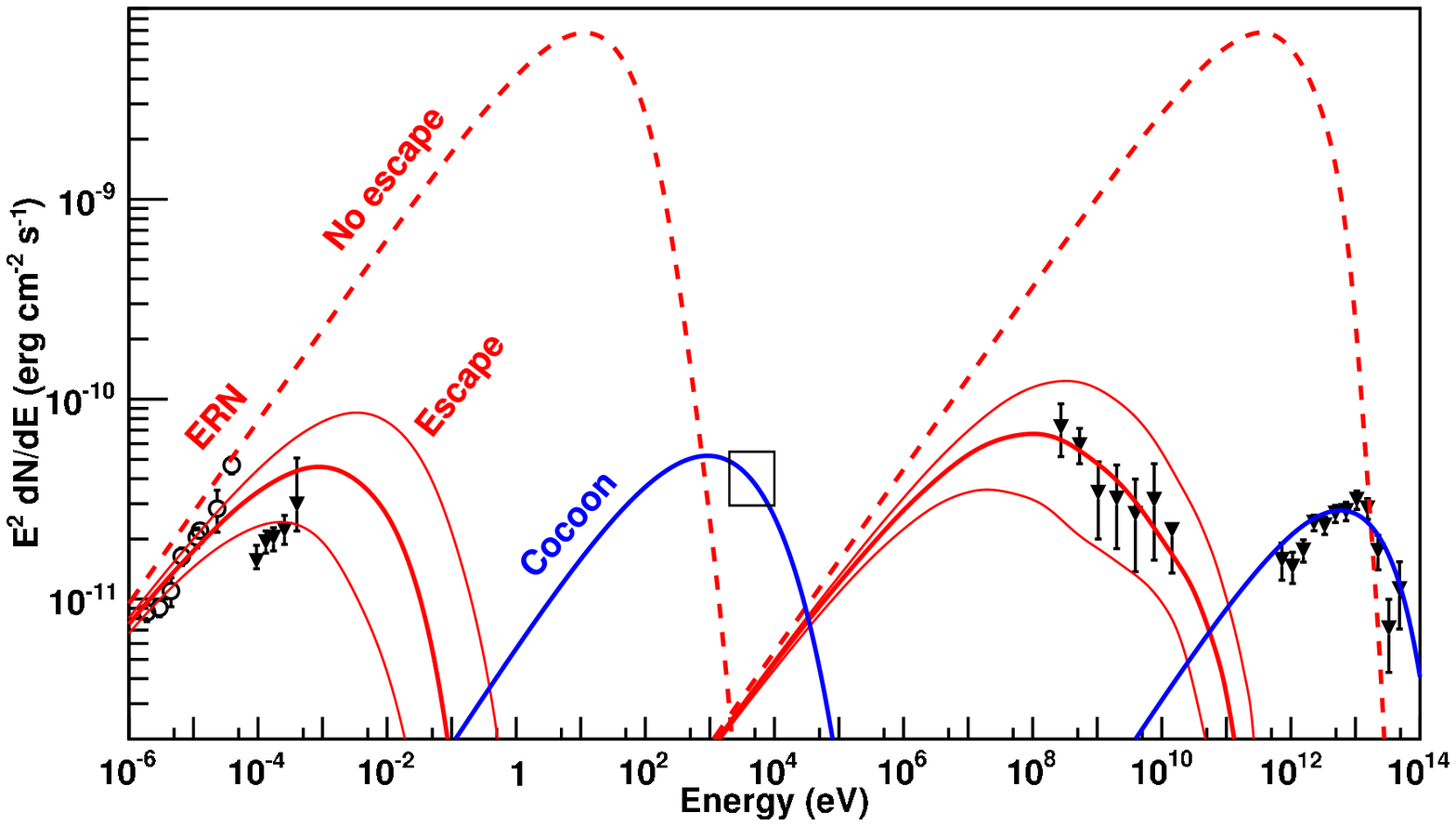}
  \caption{Pulsar wind nebula Vela-X spectral energy distribution. See~\cite{2011ApJ...743L...7H} for details.
  }
  \label{fig:VeX}
\end{figure}
Several scientific cases can be addressed by the ASTRI mini-array. For the first time, the energy range above
a few tens of TeV can be explored with an improved sensitivity compared to the current IACTs.
The nearby and powerful pulsar wind nebula (PWN) Vela-X is a typical source which can be considered 
as a primary target for the ASTRI mini-array.
Figure~\ref{fig:VeX} shows its spectral energy distribution (SED), as reported in~\cite{2011ApJ...743L...7H}, where
a clear peak is visible in the H.E.S.S. data (\cite{2006A&A...448L..43A}) at about 10~TeV, and a cut-off at about
70~TeV, making the ASTRI mini-array crucial to explore this portion of the SED.

Supernova remnants (SNR) are typical Galactic TeV emitters. RX~J1713.7$-$3946 is a young shell-like SNR
which could be considered as an excellent laboratory to investigate the cosmic ray acceleration
(see~\cite{2000A&A...354L..57M} and~\cite{2006A&A...449..223A}).
\begin{figure}
  \includegraphics[angle=0,width=.30\textheight]{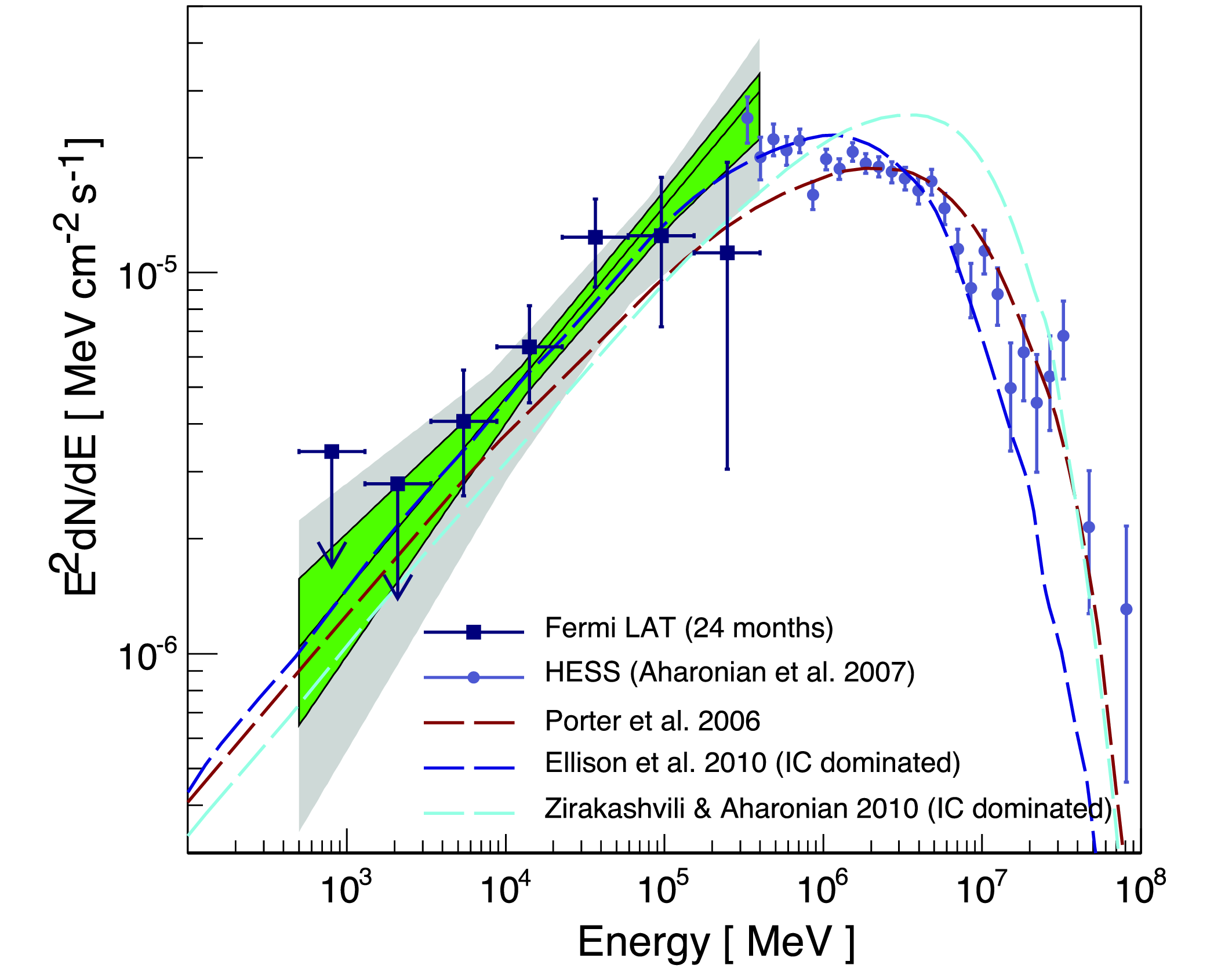}
  \caption{Supernova remnant RX~J1713.7$-$3946. See~\cite{2011ApJ...734...28A} for details.
  }
  \label{fig:171}
\end{figure}
The recent detection of this SNR by {\it Fermi} (\cite{2011ApJ...734...28A}) and the combined study with H.E.S.S.
(see Figure~\ref{fig:171}), show that the high-energy and very high-energy (VHE) emission could be interpreted in the 
framework of a leptonic scenario.
Nevertheless, the good energy resolution of the ASTRI mini-array above 10~TeV and its improved sensitivity beyond
a few tens of TeV, will improve our knowledge on the main emission mechanism acting in this source in the GeV and TeV
energy bands.
 
\begin{figure}
  \includegraphics[angle=0,width=.30\textheight]{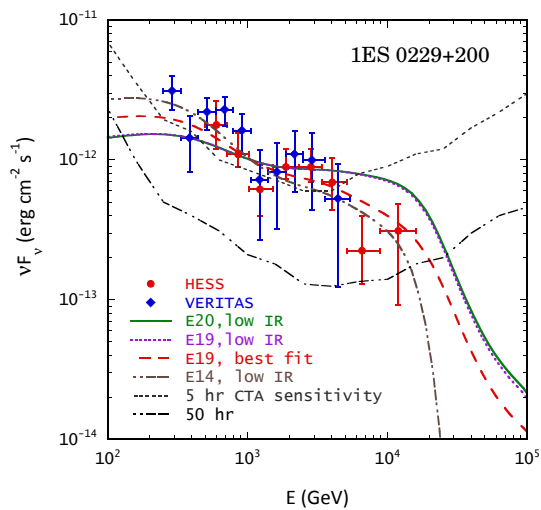}
  \caption{SED of the extreme blazar 1ES~0229$+$200 (see~\cite{2012ApJ...749...63M} for details).
  }
  \label{fig:022}
\end{figure}

The ASTRI mini-array will be extremely important to investigate the VHE emission from extragalactic sources as well.
Figure~\ref{fig:022} shows the SED of the extreme blazar 1ES~0229$+$200 (\cite{2012ApJ...749...63M}).
A clear detection of VHE emission above a few tens of TeV from such a blazar could provide fundamental
information on the long-standing debate on the emission mechanisms in this energy band. In particular,
since the cosmic-ray-induced cascade displays a significantly harder spectrum above 10--20 TeV, 
a detection above $\sim 30$\,TeV would be only compatible with an hadronic origin of the gamma-rays 
(\cite{2012ApJ...749...63M}). 

\section{SUMMMARY}
The ASTRI SST--2M end-to-end prototype will be installed and operated during Spring 2014 at the INAF Observing Station 
in Serra La Nave, Sicily. The ASTRI prototype performance will provide crucial information on several topics, such as the 
dual-mirror Schwarzschild-Couder optical design, the SiPM-based focal surface and the software/data-handling architecture, 
all of them innovative with respect to the current IACT design. Moreover, the prototype site will allow us to obtain a direct
measurement of prominent gamma-ray sources, such as the Crab nebula, MRK~421 and MRK~501. The planned ASTRI
mini-array, operated starting from 2016, will constitute the first {\it seed} of the future CTA Project, and will be open to the
CTA Consortium for both technological and scientific exploitation.

\bigskip 
\begin{acknowledgments}
This work was partially supported by the ASTRI Flagship Project financed by the Italian Ministry of Education, University, and Research (MIUR) and lead by the Italian National Institute of Astrophysics (INAF). \\
\end{acknowledgments}
\bigskip 

\hyphenation{Post-Script Sprin-ger}

\end{document}